\documentstyle[twocolumn,aps,floats,
prl,latexsym,amsmath,eqsecnum]{revtex}
\begin{document}
\author{I. Radinschi\thanks{%
iradinsc@phys.tuiasi.ro}}
\author{``Gh. Asachi'' Technical University, Iasi, 6600, Romania}
\title{Energy of a Conformal Scalar Dyon Black Hole}
\maketitle

\begin{abstract}
We obtain the energy of a conformal scalar dyon black hole (CSD) by using
the energy-momentum complexes of Tolman and M\o ller. The total
gravitational energy is given by the CSD charge in the both prescriptions.

PACS: 04. 20.-q; 04. 70.-s

\end{abstract}


\section{Introduction}

The localization of energy is a long-standing problem in the theory of
general relativity. Also, it is claimed that the energy cannot be localized.
We do not share this opinion.

It is possible to evaluate the energy and momentum distribution by using
various energy-momentum complexes. There exist an opinion that the
energy-momentum complexes are not useful to get meaningful energy
distribution in a given geometry. Virbhadra and his collaborators re-opened
the problem of the energy-momentum localization by using the energy-momentum
complexes. The results obtained for some space-times lead to the conclusion
that different energy-momentum complexes give the same energy distribution
for a given space-time [1]-[12]. Recently, Virbhadra [13] used some of these
results and investigated the Seifert conjecture. The author calculated the
energy distribution of a dilaton dyonic black hole [14] and the energy of
the Bianchi type I solution [15]. Also, we obtained the energy distribution
in a static spherically symmetric nonsingular black hole space-time [16].

The purpose of this paper is to compute the energy distribution of a
conformal scalar dyon black hole by using the Tolman and M\o ller
prescriptions. We use the geometrized units $(G=1,c=1)$ and follow the
convention that the Latin indices run from $0$ to $3$.

\section{Energy in the Tolman prescription}

Virbhadra [17] gave an exact solution of Einstein--Maxwell conformal scalar
field equations which is a black hole solution and is characterized by the
scalar charge, electric charge and magnetic charge. This solution is given
by the line element

\begin{equation}
\begin{array}{c}
ds^2=\left( 1-{\frac{Q_{CSD}}r}\right) ^2dt^2-\left( 1-{\frac{Q_{CSD}}r}%
\right) ^{-2}dr^2- \\ 
-r^2(d\theta ^2+\sin ^2\theta d\varphi ^2)
\end{array}
\end{equation}
and the conformal scalar field is given by

\begin{equation}
\psi =\sqrt{\frac 3{4\pi }}\left( \frac{q_s}{r-Q_{CSD}}\right)
\end{equation}
with $q_s$ the scalar charge, and where

\begin{equation}
Q_{CSD}=\sqrt{q_s^2+q_e^2+q_m^2}
\end{equation}

In (2.3) $q_e$ and $q_m$ are the electric charge and, respectively, the
magnetic charge.

The only non-zero components of the electromagnetic field tensor are

\begin{equation}
F_{rt}={\frac{q_e}{r^2}},\qquad F_{\theta\varphi}=q_m\sin\theta.
\end{equation}

The solution given by (2.1) has an event horizon at $r=Q_{CSD}$. Also, this
solution is the magnetic generalization of Bekenstein's solution [18].

Tolman's energy-momentum complex [19] is given by

\begin{equation}
\gamma _i^{\,\,k}={\frac 1{8\pi }}U_{i\,\,\,\,,l}^{\,\,kl},
\end{equation}
where $\gamma _0^{\,\,0}$ and $\gamma _\alpha ^{\,\,0}$ are the energy and
momentum components. We have

\begin{equation}
U_i^{\,\,kl}=\sqrt{-g}\left( -g^{pk}V_{ip}^{\,\,\,\,l}+{\frac 12}%
g_i^{\,k}g^{pm}V_{pm}^{\,\,\,\,\,\,l}\right)
\end{equation}
with 
\begin{equation}
V_{jk}^{\,\,\,\,i}=-\Gamma _{jk}^i+{\frac 12}g_j^{\,i}\Gamma _{mk}^m+{\frac 1%
2}g_k^{\,i}\Gamma _{mj}^m.
\end{equation}

Also, the energy-momentum complex $\gamma _i^{\,\,k}$ satisfies the local
conservation laws

\begin{equation}
{\frac{\partial \gamma _i^{\,\,k}}{\partial x^k}}=0.
\end{equation}

The energy and momentum in Tolman's prescription are given by

\begin{equation}
P_i=\int \hskip-7pt\int \hskip-7pt\int \gamma _i^{\,\,0}dx^1dx^2dx^3.
\end{equation}

Using the Gauss theorem we obtain

\begin{equation}
P_i={\frac 1{8\pi }}\int \hskip-7pt\int U_i^{\,\,\,0\alpha }n_\alpha dS,
\end{equation}
where $n_\alpha =\left( {\frac xr},{\frac yr},{\frac zr}\right) $ are the
components of a normal vector over an infinitesimal surface element $%
dS=r^2\sin \theta d\theta d\varphi $.

We get the expression of the energy in the Tolman prescription in the case
of a general spherically symmetric space-time that is described by the line
element

\begin{equation}
ds^2=B(r)dt^2-A(r)dr^2-D(r)r^2(d\theta ^2+\sin ^2\theta d\varphi ^2).
\end{equation}

The energy is given by 
\begin{equation}
E(r)={\frac r2}\cdot {\frac{\sqrt{A(r)B(r)}\left( A(r)-D(r)\right) }{A(r)}}.
\end{equation}

After some calculations, by using (2.12) we obtain for the CSD black hole 
\begin{equation}
E(r)=Q_{CSD}\left( 1-{\frac{Q_{CSD}}{2r}}\right) .
\end{equation}

The total gravitational energy of a CSD black hole which is obtained for $%
r\to \infty $ in (13) is equal to its CSD charge.

\section{Energy in the M\O ller prescription}

M\o ller's energy-momentum complex [20] is given by

\begin{equation}
\Theta _i^{\,\,k}={\frac 1{8\pi }}\cdot {\frac{\partial \chi _i^{\,\,kl}}{%
\partial x^l},}
\end{equation}
where 
\begin{equation}
\chi _i^{\,\,kl}=\sqrt{-g}\left( {\frac{\partial g_{in}}{\partial x^m}}-{%
\frac{\partial g_{im}}{\partial x^n}}\right) g^{km}g^{ln}.
\end{equation}

The energy in the M\o ller prescription is given by

\begin{equation}
E=\int \hskip-7pt\int \hskip-7pt\int \Theta _0^{\,\,0}dx^1dx^2dx^3={\frac 1{%
8\pi }}{\frac{\partial \chi _0^{\,\,0l}}{\partial x^l}}dx^1dx^2dx^3.
\end{equation}

The M\o ller energy-momentum complex is not necessary to carry out the
calculation in the quasi-Cartesian coordinates, so we can calculate in the
spherical coordinates.

For the line-element given by (2.1) the $\chi _0^{\,\,01}$ component is

\begin{equation}
\chi _0^{\,\,01}=\frac{2(r-Q_{CSD})Q_{CSD}\sin \theta }r
\end{equation}

Now, substituting (3.4) in (3.3) and applying the Gauss theorem we obtain
the energy of the CSD black hole

\begin{equation}
E(r)=Q_{CSD}\left( 1-{\frac{Q_{CSD}}r}\right) .
\end{equation}

In the M\o ller prescription the second term in the expression of the energy
is twice the value obtained by using the Tolman energy-momentum complex.

It is important to note that the total gravitational energy of the CSD black
hole has the same expression as in the Tolman prescription and is given by
the CSD charge.

\section{Discussion}

The main purpose of the present paper is to show that it is possible to
``solve'' the problem of the localization of energy in relativity by using
the energy-momentum complexes.

Bondi [21] sustained that a nonlocalizable form of energy is not admissible
in relativity so its location can in principle be found. Some interesting
results which have been found recently show that the several energy-momentum
complexes can give the same and acceptable result for a given space-time.
Also, in his recent paper Virbhadra [13] emphasized that though the
energy-momentum complexes are non-tensors under general coordinate
transformations, the local conservation laws with them hold in all
coordinate systems. Chang, Nester and Chen [22] showed that the
energy-momentum complexes are actually quasilocal and legitimate expressions
for the energy-momentum.

We have calculated the energy of a CSD black hole by using the Tolman and
M\o ller energy-momentum complexes. In the Tolman prescription the energy
associated with the CSD black hole is found to be the same as it was earlier
evaluated by Virbhadra in the Weinberg prescription. The M\o ller
energy-momentum complex gives for the second term in the expression of the
energy twice the value obtained by using the Tolman energy-momentum complex.
The total gravitational energy of the CSD black hole obtained when $r\to
\infty $ is the same in the both Tolman and M\o ller prescriptions and is
equal to the CSD charge.

Also, we obtain the expression of the energy distribution in the case of a
general static spherically symmetric space-time by using the Tolman
prescription.

\section{Acknowledgments}

I am grateful to K. S. Virbhadra for his helpful advice.

\end{document}